\documentclass[conference]{IEEEtran}
\IEEEoverridecommandlockouts
\usepackage{cite}
\usepackage{amsmath,amssymb,amsfonts}
\usepackage{algorithmic}
\usepackage{graphicx}
\usepackage{textcomp}
\usepackage{xcolor}
\def\BibTeX{{\rm B\kern-.05em{\sc i\kern-.025em b}\kern-.08em
    T\kern-.1667em\lower.7ex\hbox{E}\kern-.125emX}}

\newcommand{\sectopic}[1]{\vspace{0.2em}\par\noindent{\textit{\bfseries #1}}}
    
\begin{document}

\title{Multi-Modal Emotion Recognition for Enhanced Requirements Engineering: A Novel Approach}

\author{
   \IEEEauthorblockN{Ben Cheng\IEEEauthorrefmark{1}, 
   Chetan Arora\IEEEauthorrefmark{2},
   Xiao Liu\IEEEauthorrefmark{1},
   Thuong Hoang\IEEEauthorrefmark{1},
   Yi Wang\IEEEauthorrefmark{1},
   John Grundy\IEEEauthorrefmark{2}   
   }
   \IEEEauthorblockA{\IEEEauthorrefmark{1}School of Information Technology, Geelong, Victoria, Australia
  \\\{chengye, xiao.liu, thuong.hoang, xve\}@deakin.edu.au}
  \IEEEauthorblockA{\IEEEauthorrefmark{2}Faculty of Information Technology, Monash University, Clayton, Victoria, Australia
   \\\{chetan.arora, john.grundy\}@monash.edu}

}


\maketitle
\thispagestyle{plain}
\pagestyle{plain}
\pagenumbering{arabic}

\begin{abstract}
Requirements engineering (RE) plays a crucial role in developing software systems by bridging the gap between stakeholders' needs and system specifications. However, effective communication and elicitation of stakeholder requirements can be challenging, as traditional RE methods often overlook emotional cues. This paper introduces a multi-modal emotion recognition platform (MEmoRE) to enhance the requirements engineering process by capturing and analyzing the emotional cues of stakeholders in real-time.
MEmoRE leverages state-of-the-art emotion recognition techniques, integrating facial expression, vocal intonation, and textual sentiment analysis to comprehensively understand stakeholder emotions. This multi-modal approach ensures the accurate and timely detection of emotional cues, enabling requirements engineers to tailor their elicitation strategies and improve overall communication with stakeholders. We further intend to employ our platform for later RE stages, such as requirements reviews and usability testing. By integrating multi-modal emotion recognition into requirements engineering, we aim to pave the way for more empathetic, effective, and successful software development processes. We performed a preliminary evaluation of our platform. This paper reports on the platform design, preliminary evaluation, and future development plan as an ongoing project. 
\end{abstract}


\section{Introduction}~\label{sec:introduction}

Requirements Engineering (RE) is a critical phase in the software development lifecycle, responsible for capturing, documenting, and managing stakeholders' needs to create a set of well-defined and comprehensive system requirements \cite{aurum2005}. The effectiveness of RE directly impacts the quality and success of the resulting software system. The RE process often involves complex human interactions, which in turn involve the emotions of all stakeholders. The emotional cues, if overlooked in RE, can often lead to misunderstandings and misinterpretations~\cite{ramos2005emotion,colomo2010study}.

Emotions are significant in human communication, influencing decision-making, problem-solving, and collaboration~\cite{isen2001influence,colomo2010study}. In the context of RE, understanding stakeholders' emotions can provide valuable insights into their needs, priorities, and expectations, ultimately leading to more effective communication and elicitation of requirements. Despite its importance, emotional awareness has not been extensively explored in the field of RE or software engineering in general~\cite{sanchez2019taking}.

Recent advances in Artificial Intelligence (AI) and machine learning have led to significant improvements in emotion recognition techniques in videos, audios, and texts~\cite{hazarika2020, zamani2021machine, DAS23}. These advancements have opened new opportunities to integrate emotion awareness into RE processes. However, there are also many practical challenges in using emotion recognition techniques in RE processes. For example, how to achieve real-time emotion recognition in various RE environments; how to generate useful emotion analysis result when stakeholders are experiencing intensive emotional changes; and how to protect user privacy since many personal videos and audios are being used by emotion recognition models. Therefore, the application of emotion recognition in RE processes is still in its infancy and there are many open issues to be addressed. This paper presents a novel multi-modal emotion recognition platform (MEmoRE) to enhance the RE process by capturing and analyzing emotional cues in stakeholder interactions. We plan to leverage MEmoRE in different RE activities, including stakeholder interviews (elicitation), change management (specification and validation), requirements review meetings (validation), and usability testing (validation).

The primary objectives of this paper are:
\begin{enumerate}
\item To introduce MEmoRE - a multi-modal emotion recognition platform that integrates facial expression, vocal intonation, and textual sentiment analysis to comprehensively understand stakeholder emotions during RE.

\item To discuss the potential benefits and challenges associated with incorporating emotion awareness into requirements engineering, emphasizing its impact on communication, elicitation strategies, and overall project success.

\item To present a preliminary evaluation of the proposed platform, highlighting its effectiveness in a real-world scenario and outlining a future evaluation plan as part of an ongoing project.
\end{enumerate}

\sectopic{Structure.} The remainder of this paper is structured as follows: Section~\ref{sec:relatedwork} reviews the related work. Section~\ref{sec:overview} presents the platform overview and its key components. Section~\ref{sec:results} presents the details of the preliminary evaluation results, discussions, and current limitations. Section~\ref{sec:roadmap} discusses the research roadmap, and Section~\ref{sec:conclusion} concludes the paper.

\section{Related Work}\label{sec:relatedwork}

\sectopic{Emotions in RE.} Requirements engineering (RE) focuses on capturing and analyzing user requirements, such as user feedback data. While user data often reflect user behavior and experience, there is a notable lack of research exploring the relationship between user behavior, experience, and emotions~\cite{stade19}. In recent years, some studies have explored human emotions' role in the RE process, such as requirements changes \cite{Madampe23, Mohammed-Amr22} and sentiment analysis of user reviews~\cite{guzman2014users,bano2021rise}. While these studies have presented the connection between human emotions and RE, emotions have not been extensively considered in the software development lifecycle. If emotions are not adequately considered, this can lead to misunderstandings of user needs, user dissatisfaction with the software product, and even uninstallation of the application~\cite{stade19}. Emotion-oriented requirements engineering aims to explore the impact of human emotions in various application scenarios of RE activities \cite{Maheswaree19}.
This paper presents the application of a multi-modal emotion recognition approach to enhance the accuracy of analyzing user emotional results and understanding the relationship between emotions and user requirements in RE activities.

\sectopic{Detecting emotions.} Measuring emotions primarily includes physiological data, behavioral data, facial data, and self-report data \cite{meiselman2016}. For example, to address deliberately fabricated smiles, Monica et al. \cite{Perusqu19} employed computer vision (CV) and facial distal electrocardiography (EMG) to identify different types of user smiles. The study ultimately demonstrated that CV could recognize user behavior that is not discernible by humans, while EMG can detect concealed behaviors that are not visible. However, most studies currently focus on the influence of a single data source on user requirements. Single-source emotion data frequently produce imprecise results.
Nevertheless, a recent study focused on applying multi-source emotion recognition in RE activities \cite{cheng21}. Cheng et al. \cite{cheng21} reported that single-source emotion recognition is insufficient for providing comprehensive emotional analysis, which may hinder engineers' understanding of user emotions. They presented an edge computing-based multi-source emotion recognition platform (Edge4Emotion) that can analyze user emotions from various angles, such as facial expressions and audio data. Meanwhile, some research focuses solely on multi-modal emotion frameworks, models, and algorithms \cite{hazarika2020, mittal2020}. Our approach to multi-modal emotion recognition is novel and can substantially benefit the RE community.

\sectopic{Applications.} Currently, several studies mainly focus on evaluating the relationship between emotions and specific user groups (e.g., elderly) \cite{Demaeght22, Boateng21}, teams (e.g., agile teams) \cite{Mohammed-Amr22}, and application scenarios (e.g., smart home systems, video games, and robot assistance) \cite{Maheswaree19, Demaeght22, Callele08}. Human emotions are key in accepting software applications, especially in social systems. For example, Miller et al. \cite{MILLER2015} reported that adding the concept of emotional goals improved emergency systems, resulting in higher user satisfaction compared to earlier systems. Curumsing et al. \cite{Maheswaree19} also reported an evaluation example of employing an emotion-oriented approach to evaluate the SofiHub application. They found that the SofiHub application effectively mitigated the loneliness experienced by older users and made them feel safer and more cared for.
Furthermore, emotions have a significant impact on software developers. For example, agile teams have significant emotional effects on developers during requirement changes. However, currently, no effective tools are available for detecting and analyzing developers’ emotions \cite{Kołakowska13, Mohammed-Amr22}. Therefore, there is a need for tools and techniques that can accurately detect and analyze emotions in the context of software development, which can improve the effectiveness and efficiency of software development and enhance user satisfaction.

In summary, the application of multi-modal emotion recognition has been unexplored in RE activities. Therefore, our study will be an important step in the direction of applying multi-modal emotion recognition in RE activities.

\section{Multi-modal Emotion Recognition Platform}\label{sec:overview}

In this section, we describe our multi-modal emotion recognition platform for RE (MEmoRE), the main components developed and planned, and the planned applications in RE.

\subsection{System Overview}
\begin{figure}[!t]
\centering
\includegraphics[width=0.95\linewidth]{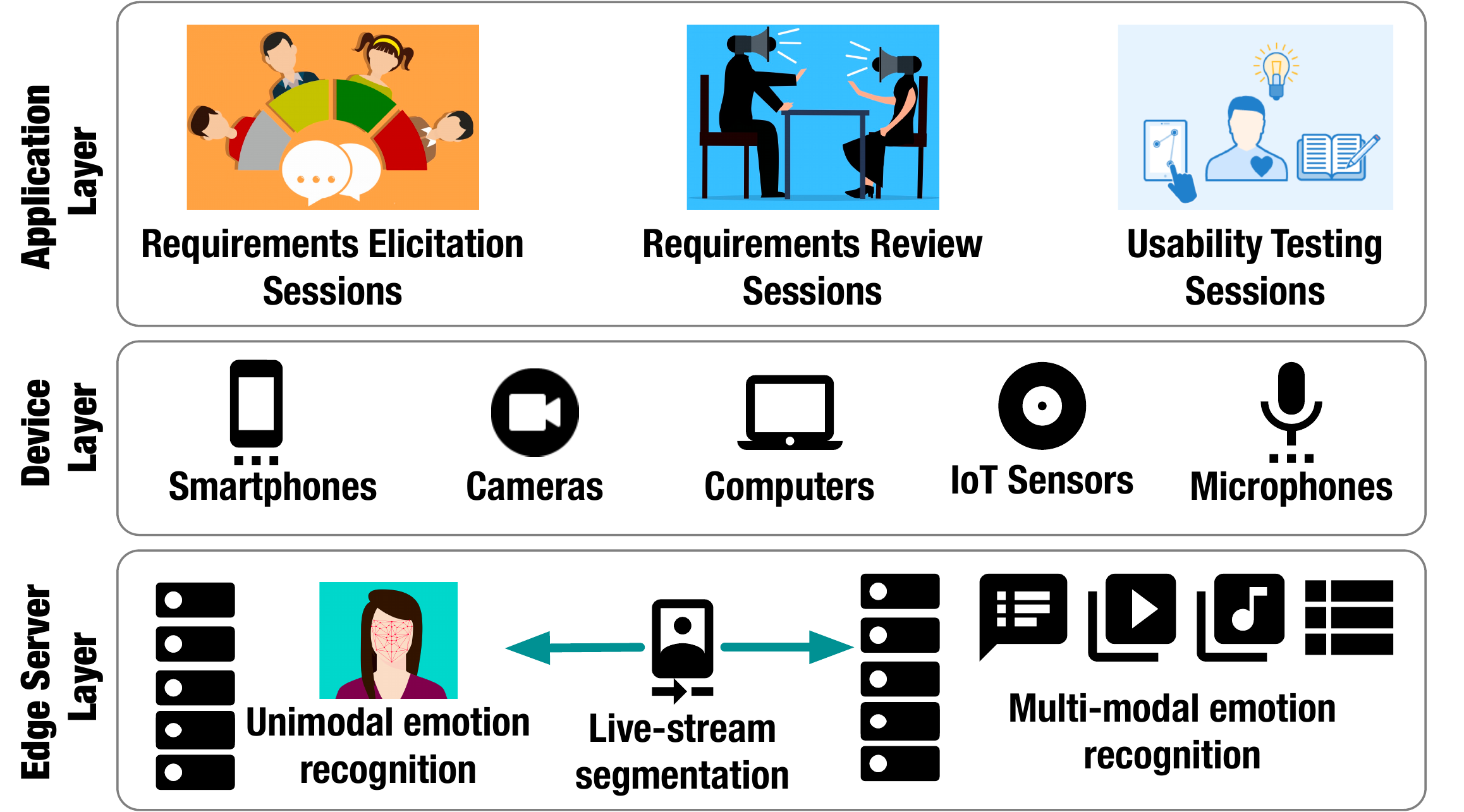}
\caption{MEmoRE Platform Overview }
\label{fig:overview}
\vspace*{-1em}
\end{figure}
The architecture of the system is illustrated in Figure~\ref{fig:overview}. It can be categorized into three distinct layers: the application, the device and the edge server layers.

The device layer encompasses a variety of devices such as cameras, smartphones, microphones and  wearable devices, which can gather motion data and physiological data from multiple sources. These devices can connect to the edge servers through wired networks and wireless connections. The edge server layer comprises multiple edge servers capable of processing various types of data and supporting different emotion recognition models for uni-modal and multi-modal. For instance, the platform includes dedicated edge servers for recognizing emotions based on video and audio data separately. Moreover, an important edge server for multi-modal emotion recognition processes both visual and audio data simultaneously. The application layer consists of various tools and applications catering to different RE activities. For example, the emotion recognition results for interviewees can be utilized in requirements gathering to understand their genuine opinions about specific requirements, while the results for testers in usability testing can help identify their actual experiences with certain functionalities. Further details about the various components of the MEmoRE platform are provided below.
\begin{figure}[!t]
  \centering
  \includegraphics[width=\linewidth]{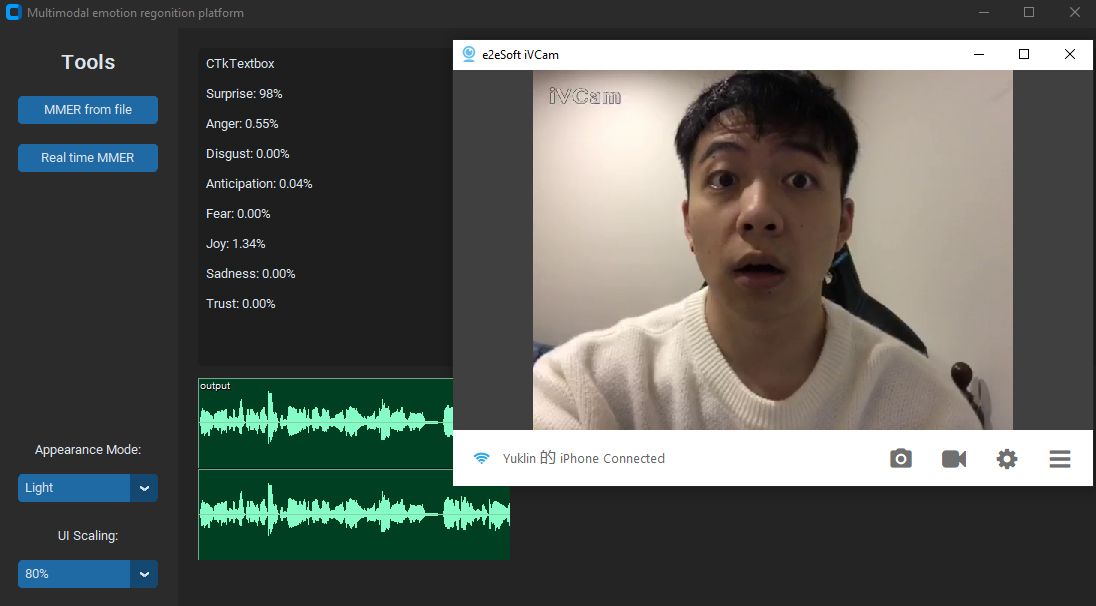}
  \caption{Example UI and emotion recognition accuracy for video segments of various lengths.}
   \label{fig:example}
   \vspace*{-2em}
\end{figure}
An example UI of the platform is shown in Fig.~\ref{fig:example} which also demonstrates the accuracy of emotion recognition for video segments with various lengths. The process can be described as follows:

\begin{enumerate}

\item The end devices will capture the user's physiological signals which constantly change during a conversation. At the moment, for our multi-modal emotion analysis, we focus on facial expressions and vocal cues, so we will use cameras and microphones to capture the user's facial expressions and vocal changes during the conversation.

\item The data captured by the camera and microphone will be pre-processed by the video processing edge server. The video processing edge server will segment the data stream (e.g., every 10-seconds/30-seconds) and save these vidoe segments in the MP4 video format. This facilitates data recording and manual review.

\item The edge server for multi-modal emotion recognition will extract features from the video and audio data separately, which will then be fed into a pre-trained deep learning model. Ultimately, the predicted emotional state will be generated as the output for further usage by various RE activities.

\item Our approach involves incorporating the output results into our requirements analysis. This enables us to identify the emotional state of the target user during the requirements gathering or testing phase, helping software engineers to focus on the users' specific requirements and improve the design accordingly.
\end{enumerate}

\subsection{Major Components}

\sectopic{Application Layer.} MEmoRE platform can support different RE activities, including requirements elicitation, validation, and so on. Requirements elicitation is an important step for stakeholders in the software development process. However, traditional requirements elicitation techniques can limit their ability to dig into stakeholders' true needs and desires. Using the emotion recognition approach enhances the requirements elicitation process, which better feeds the expectation of stakeholders. With our multi-modal emotion recognition platform, we can easily analyze stakeholders' emotions in real-time discussion environments to identify important requirements. When stakeholders express high positive emotions during a discussion, it might indicate that particular requirements are essential to them. In contrast, if stakeholders exhibit confusion or frustration, it often suggests that our engineers may not have understood their requirements correctly. We need to reconsider the proposed solution or clarify the requirements further. In this way, our platform can help facilitate more effective communication and collaboration between stakeholders and engineers, leading to the identification of requirements that better address the needs and desires of stakeholders. Furthermore, emotion recognition can prioritize requirements based on their emotional impact. For instance, requirements related to user satisfaction or emotions such as joy, trust, or confidence can be given higher priority than others. This prioritization can ensure that requirements with top importance are given priority during the development process.

Requirements validation is essential to ensure that they accurately reflect the stakeholders' needs and expectations. This can be done by observing stakeholders' emotional responses to prototypes or demos. It is a common practice to provide prototypes to stakeholders which allows engineers to gather feedback on the design, functionality, and overall user experience. During the demonstration or the usage of the prototype, stakeholders may reveal various emotional responses. For example, positive emotions such as surprise and joy indicate that the software's functionality has met or exceeded the expectation of users. Conversely, if a user expresses disgust or frustration, it may indicate that the software does not meet the user's expectations or is difficult to use. Our platform can automate this process and generate reports that provide engineers with detailed records of emotional responses to specific requirements so that engineers can adjust the development direction and improve the user experience. Furthermore, emotion-based requirement validation can be integrated into the requirement engineering process to enhance the quality of requirements and ensure their alignment with stakeholders' emotional needs and preferences.

\sectopic{Device Layer.} The end device layer in our platform supports the connection of various types of devices and sensors for emotion data collection to collect real-time emotion data. The data is then transmitted to different edge servers for emotion recognition. Cameras provide live-stream video data and microphones provide audio data which supports different edge servers to perform emotion recognition.
Moreover, smartphones can serve as integrated devices that facilitate the collection of multi-modal data, such as video and audio, through embedded cameras and microphones. Overall, the end device layer is an essential component of our platform as it allows for the integration of various devices for emotion data collection. By utilizing a combination of different sensors and devices, our platform can collect multi-modal data to provide a more accurate, reliable and comprehensive emotion analysis.

\sectopic{Edge Server Layer.} This layer consists of various computing nodes near the network edge (namely edge servers) for processing data transmitted directly from the end devices. Our novel approach focuses on the MEmoRE server to provide emotion fusion results from human facial expression, human voice and text. Specifically, it first segments the real-time video and audio streams into small clips with fixed lengths as input for a pre-trained multi-modal emotion recognition model, and then outputs the probabilities for eight different emotions as classified in \textcolor{black}{Joy, Sadness, Anger, Anticipation, Disgust, Fear and Trust.} For multi-modal emotion recognition in our platform, we utilized the model proposed in \cite{zhao2020end} which employ a deep Visual Audio Attention Network (VAANet). VAANet is a novel architecture that integrates spatial, channel-wise, and temporal attentions into a visual 3D convolutional neural network (CNN) and temporal attention into an audio 2D CNN. In addition, our platform also supports uni-modal emotion analysis, specifically for image and audio modalities. Each of these modalities can be handled by a dedicated edge server that utilizes deep learning models for emotion recognition. In fact, applying different uni-modal emotion recognition methods under specific requirements engineering environments can be beneficial. For example, if the environment involves a lot of phone conversations, using an audio-based emotion recognition method would be more appropriate. In contrast, if the environment involves video conferencing or surveillance footage, a video-based emotion recognition method would be more useful. Furthermore, incorporating text-based emotion recognition methods could be more useful for requirements engineering environments that involve a lot of written communication, such as email or chat-based communications.

\section{Preliminary Evaluation}\label{sec:results}

\subsection{Preliminary Results}

Here, we demonstrate the preliminary results of the experimentation on MEmoRE. We divided the experiment into two steps. First, we aimed to determine the optimal length of segments, in seconds, for achieving the highest accuracy when segmenting the interview data. Second, we aimed to evaluate the emotion results of the interview data, using MEmoRE. In the following, we describe the details of the experiments.

\sectopic{(1) Experiment 1: } We assessed the accuracy of different video segment lengths using eight minute long interview videos. The video format was MP4. To ensure uniformity of the interview video data, we set the frame rate to 24 frames per second (24 fps), which is a common video standard. Next, we segmented the video into different intervals of 6-seconds, 10-seconds, 15-seconds, 30-seconds and 60-seconds. Fig.~\ref{fig:fig4} shows the accuracy results for videos segmented into different lengths. We found that the lowest accuracy was achieved when then interview video data was segmented into 60 seconds. If the interview video data was segmented into 6-seconds, 15-seconds and 30-seconds, the accuracy was below 50\%. After multiple validations, we discovered that the highest accuracy was achieved only when the video was segmented into 10 seconds. With longer videos, e.g., 60 seconds, the platform is inaccurate potentially due to multiple emotions in the video and since the platform is currently tuned to provide a single classification. With shorter videos, there is potentially insufficient content to accurately judge emotions.

\begin{figure}[!t]
  \centering
  \includegraphics[width=0.8\linewidth]{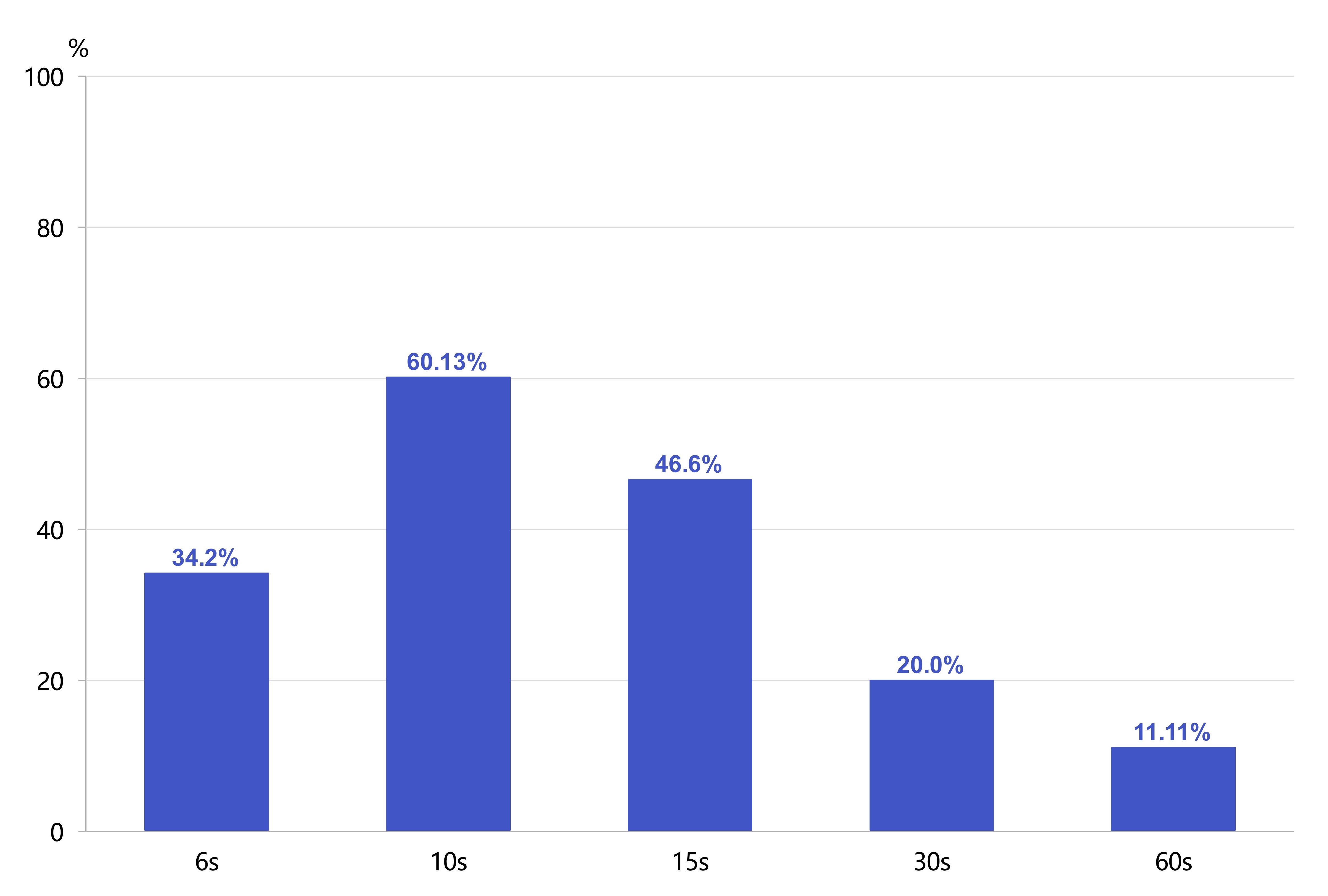}
  \caption{Accuracy of emotion recognition at different segment lengths.}
   \label{fig:fig3}
\end{figure}

\begin{figure}[!t]
  \centering
  \includegraphics[width=0.8\linewidth]{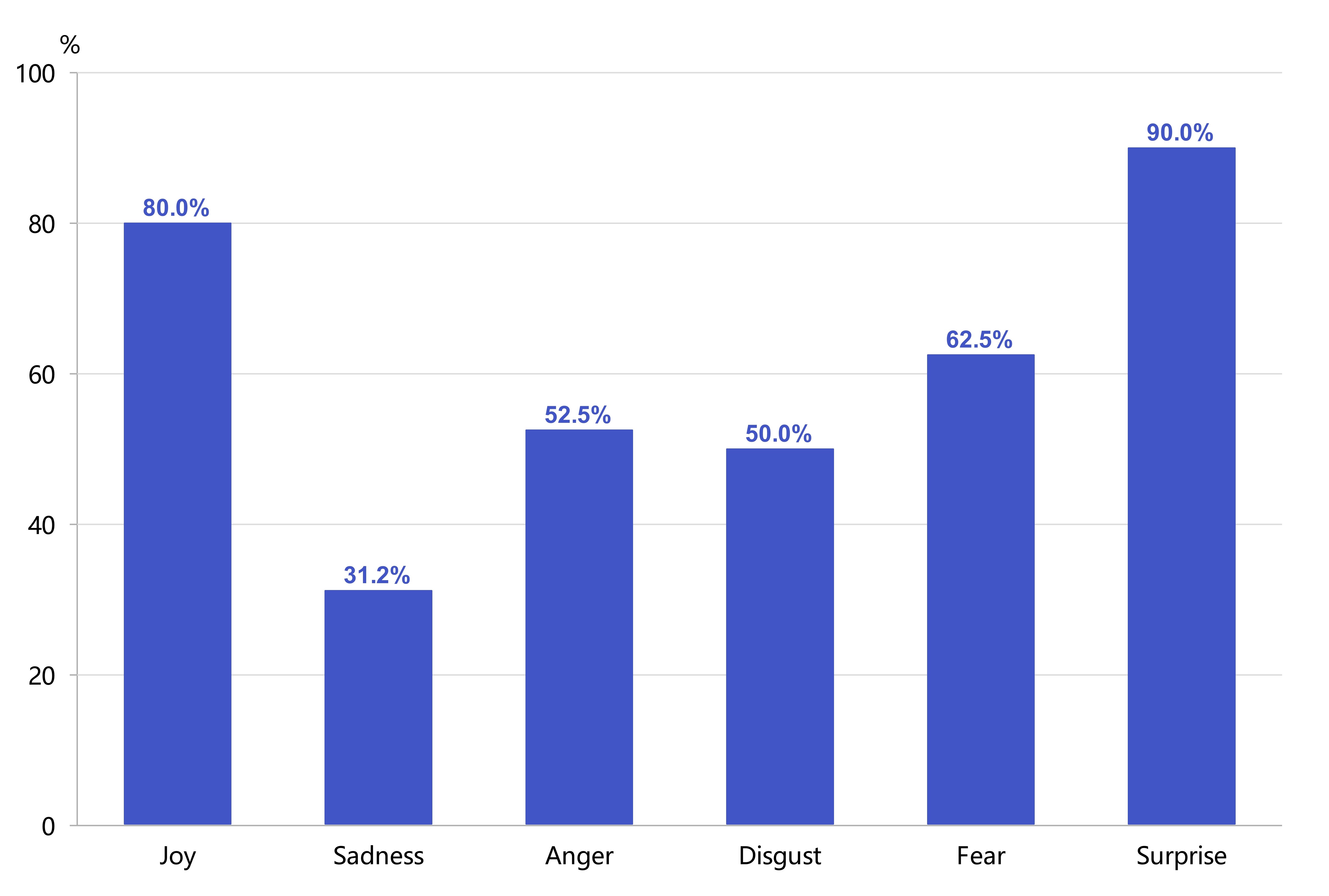}
  \caption{Accuracy of emotion detection for  different emotions}
   \label{fig:fig4}
\end{figure}

\sectopic{(2) Experiment 2:} We aimed to discover the model's performance against different emotions. In this experiment, we selected 48 videos with six different emotions and eight videos for each type of emotion from the MELD dataset~\cite{poria2018meld}. Fig.~\ref{fig:fig4}  shows the accuracy results of the model recognizing different emotions. The results showed that the accuracy of detecting positive emotions is high compared to negative emotions. For instance, the model performs well for detecting joy and surprise, whereas, not accurately at detecting sadness, anger and disgust. We plan to train further and improve our platform on more data and improve the accuracy for all emotions.

\subsection{Discussion}\label{sec:discussion}

\sectopic{The Importance of Video Segmentation.} It is well known that human emotions play a significant role in requirements engineering activities. This also means that low accuracy in emotion recognition could result in incorrect decisions and \textcolor{black}{analysis} by engineers. In this work, it was observed that the highest accuracy is achieved using a 10-second interval for real-time video segmentation. However, this could be varied under different RE environments. Hence, we suggest to test and to determine the best video segmentation interval to improve the accuracy of real-time emotion analysis results. This may assist requirement engineers in capturing complete conversations and better understanding end-users needs. Moreover, it was noted that using a 6-second interval might be too short for some conversations and reactions to complete. It is very challenging to accurately detect human emotions using a very short segment with limited information. For example, a person may cry out of extreme joy instead of sadness, but this emotion can be easily misunderstood if the video segment is very short. However, using much longer segments such as 30-second and 60-second intervals might also not yield accurate results as the conversation content and user's emotional state can change multiple times, making it difficult to determine the true emotion and hence less useful for RE activities.




\sectopic{The Importance of Emotion Recognition Accuracy in RE Activities.} Human emotions present a complex process in RE activities and the environments for various RE activities could also be very different. Therefore, relying on single-source emotional analysis may be insufficient to accurately capture users' emotions. In this work, we propose the MEmoRE platform that identifies users' emotions through  multi-modal combination of facial expressions, vocal intonation, and textual sentiment. The results of the second experiment demonstrated (Figure \ref{fig:fig3}) that the highest accuracy rates are observed for surprise and joy in the real-time interview video. In addition, the MEmoRE platform also detected some negative emotions, such as disgust, fear, and sadness. These negative emotions are also important in RE activities as they can help engineers identify whether stakeholders have negative emotions, and if so, engineers can further investigate the cause for them.

In summary, here we present some potential contributions of the MEmoRE platform to the RE community as follows:

\begin{enumerate}
    \item Multi-modal emotion recognition can assist requirement engineers in validating requirements, gauging stakeholder engagement, and ensuring that the software product meets stakeholders' expectations~\cite{ferrari2021using}.
    
    \item By identifying users' positive and negative emotions, requirement engineers can gain a better understanding of how the software performs and where needs improvement.
    \item Our MEmoRE platform can automate the process for emotion detection and analysis. It can assist requirement engineers in efficiently determining requirement priorities based on stakeholders' emotional responses.
    \item Stakeholders' emotional responses can guide requirements engineers in selecting more effective elicitation methods, and help them better understanding stakeholders' needs and expectations. 
\end{enumerate}

\subsection{Current Limitations}

There are still some limitations in our current work. Although our research indicates that segmenting every 10~seconds produces better results, in actual RE activities, the conversation interval is not always fixed. It is possible to switch between two or more emotions in seconds, increasing the probability of errors. 
Furthermore, the current design of MEmoRE includes text, video, and audio modalities. There is still bias due to individual differences in facial expressions. To address this issue, we plan to incorporate more modalities.
Moreover, we have not tested our platform for actual RE activities in the presented experiments. To address this, we plan to conduct user studies with university students, as detailed in the following section \ref{sec:roadmap}. 

\section{Research Roadmap}\label{sec:roadmap}

\subsection{Evaluation Plan}

\sectopic{Recruit Participants.} We aim to recruit participants in information technology, computer science and design courses who have completed or are current students in tertiary education. To ensure valuable data from participants' feedback, we require that they are currently in the final year of their tertiary courses. We plan to recruit $\approx$25 students for this experiment. We further plan to conduct case studies with practitioners (from our industry partners) in healthcare, e-commerce and app development domains for different RE tasks, e.g., elicitation~\cite{aurum2005}, negotiations~\cite{ferrari2021using}, change management~\cite{arora2015narcia}.


\sectopic{Experimental Design.} We plan to deploy the experiment in software projects of student teams. The experiment time will be $\approx$45 minutes. We will ask the participant groups to discuss their team software project progress in $\approx$15 minutes. We plan to collect and analyze the participants' emotions during this time. After the discussion, we will give the participants their emotion recognition results during the interviews in $\approx$10 minutes. Participants can discuss these results together. For example, if some participants were confused during the interviews, they could focus on the negative emotion in the last 15 minutes. After completing the experiment, participants must fill out a five-point Likert scale questionnaire. The questionnaire aims to collect feedback on multi-modal emotion recognition's positive and negative aspects. Throughout the experiment process, we will record video and audio data for analysis and ensure the confidentiality of user data. Regarding practitioner studies, we aim to analyze recorded requirements elicitation and review sessions and perform a retrospective analysis of MEmoRE results with practitioners. This will focus on assessing the accuracy of MEmoRE and implications of real-time emotion recognition.

\subsection{Future Work}
We plan to adopt a conversational segmentation method instead of the 10-second segmentation to address the limitations. This involves segmenting based on sentences, which allows for a more holistic recognition of an individual's emotions and addresses the limitation of the time-based segmentation method.
As ongoing work, the platform will be used as an open-source project to host future studies. Here, we also provide research directions in the near future. 
\begin{enumerate}
    \item \textbf{Using MEmoRE to support requirements changes.} Requirements changes represent a significant challenge in software development, especially agile development. Based on multi-modal emotion recognition, we can explore users’ real emotions, thus, determining their true needs and feelings. 
    \item \textbf{Expanding the platform to different methods.} There are other requirements elicitation methods besides interviews~\cite{karras2016supporting}. For example, using virtual reality (VR) to capture user needs; however, there is no prior work on stakeholder needs and emotions based on VR~\cite{wang2021virtual}. Therefore, we plan to deploy MEmoRE within a VR environment and explore user emotions and needs. 
    
    \item \textbf{Supporting special user groups and different types of application scenarios.} In the future, we plan to focus on the emotions of special user groups such as elderly, children and people with disabilities (e.g., speech impaired users). We also plan to support deployment in various scenarios, such as requirements elicitation workshops with various number of stakeholders. Overall, our platform will cover special user groups in different types of application scenarios.
\end{enumerate}
\section{Conclusion}\label{sec:conclusion}

Human emotions play a significant role in RE activities, as they are an important factor in effective communication and eliciting stakeholder needs. However, the application of emotion recognition in the RE process is still in its infancy despite the success of many deep learning based emotion recognition methods. In this paper, we presented the details of our MEmoRE platform which can assist requirements engineers in better understanding user responses towards software requirements through accurate detection of user emotions. Experimental results demonstrated that our MEmoRE platform can accurately analyze real-time interview video data and using a 10-second segmentation interval to achieve the highest accuracy. We also discussed the contributions and limitations of our MEmoRE platform for RE activities. Finally, as an ongoing research project, we have designed an evaluation plan and proposed a series of future work.


\bibliographystyle{IEEEtran}

\bibliography{ref}

\end{document}